# New Weighted Sum of Gray Gases (WSGG) Models for Radiation Calculation in Carbon Capture Simulations: Evaluation and Different Implementation Techniques

O. A. Marzouk[1,2] and E. D. Huckaby[1]

[1] U.S. Department of Energy, National Energy Technology Laboratory,
Morgantown, West Virginia 26507-0880, USA
[2] West Virginia University Research Corporation
Morgantown, West Virginia 26507-0880, USA

We apply several weighted sum of gray gases models (WSGGMs) to calculate the radiative absorption coefficient for gas mixtures containing $H_2O$ and $CO_2$. Our main objectives are to analyze and compare four WSGGMs which have been recently developed for oxy-fuel combustion. The models are compared with the widely-used air-fuel WSGGM of Smith et al. In addition to direct comparison of the absorption coefficients, we compare finite-volume solutions of the radiative equation of transfer in a 2m x 2m x 4m box with a specified inhomogeneous temperature field and a homogeneous mixture of the $H_2O$, $CO_2$ and $N_2$. Calculations using a spectral line-based WSGGM (SLW) are used as a reference solution to estimate the accuracy. For each WSGGM, we apply two interpolation methods for determining the model coefficients at arbitrary $H_2O$-to-$CO_2$ ratios. For wet-recycle oxy-fuel combustion, we found that the deviation of the air-fuel WSGGM was not significantly larger than several of the newer models. However, with dry recycle (90 vol%-$CO_2$) the air-fuel WSGGM model underpredicts the radiative flux and radiative heat source in contrast to the other models which overpredict these fields. Piecewise linear interpolation consistently improves the predictions of the air-fuel WSGGM, but only has a modest effect on the predictions of the oxy-fuel WSGGM's.

## 1 Introduction

Oxy-fuel combustion, where fossil-fuel is burned in an atmosphere free from $N_2$, is a promising technology for reducing $CO_2$ emissions from power boilers [1]. The utilization of computational modeling for developing and improving oxy-fuel test systems has been demonstrated [2–4] and is expected to be useful for the development of both new systems and retrofits. Radiation is an important mode of heat transfer in the operation of a boiler, with greater importance for oxy-firing. This is due to higher gas emissivities in oxy-firing from the higher concentration of tri-atomic gases ($H_2O$ and $CO_2$) in the furnace [5]. Consequently, the accurate calculation of the radiative absorption coefficient of the gas mixture has a greater effect on the overall solution accuracy. In addition, the ratio of $H_2O$ to $CO_2$ in oxy-fuel combustion is different than what is found in air-fuel combustion, particularly when flue gas is dried prior to recycling.

CFD (computational fluid dynamics) practitioners who utilize an air-fuel WSGGM in oxy-fuel simulations should have concerns about the decrease in overall solution accuracy caused by utilizing a radiative property sub-model outside the calibrated operating regime. In this paper, we begin to address these concerns, by analyzing the predictions from four recently (2010) developed WSGGMs [6–8] and the classical model of Smith, Shen and Friedman [9]. The model of Smith et al. was selected because it is one of the most widely-used classical WSGGMs [10] and it is incorporated in the default WSGGM for the commercial CFD software ANSYS-FLUENT [11]. The model (or Fluent) has also been used in different radiation studies [7, 12, 13]. The radiative absorption coefficient and emissivity calculated using these models are compared for a range of $CO_2$-to-$H_2O$ ratios at pathlengths of 1 m and 10 m. We also estimate the error in the radiative flux when using the different models to calculate the absorption coefficients in the solution of the radiative transfer equation (RTE). The reference solution is provided by the spectral line-based weighted sum of gray gases (SLW) model.

We use a 2m2m4m rectangular box [14] with an inhomogeneous temperature field and uniform black walls at 300 K and uniform gas composition as a test configuration. The temperature field, shown in Figure 1, is axisymmetric within an imaginary inscribed 2-m diameter cylinder given by

$$T(Z, r) = 800 + [T_{centerline}(Z) - 800] \ f(r) \tag{1}$$

where $Z$ and $r$ are the axial and radial coordinates (in m), and

$$T_{centerline}(Z) = 400 + 1400 \ \frac{Z}{0.375} \qquad \text{for } 0 \leq Z \leq 0.375 \tag{2a}$$

$$T_{centerline}(Z) = 1800 - \frac{1000}{3.625} \ (Z - 0.375) \qquad \text{for } 0.375 \leq Z \leq 4 \tag{2b}$$

$$f(r) = 1 - 3 \ r^2 + 2 \ r^3 \tag{3}$$

The portion of the box outside the imaginary inscribed cylinder has a uniform temperature of 800 K. For this temperature field, we consider four different uniform gas mixtures, which are shown in Table 1. The first and second test cases are representative of air-fuel combustion, whereas the third is representative of oxy-fuel combustion with wet recycle of flue gas, and the last is representative of oxy-fuel combustion with dry recycle of flue gas. The total pressure is 1 atm.

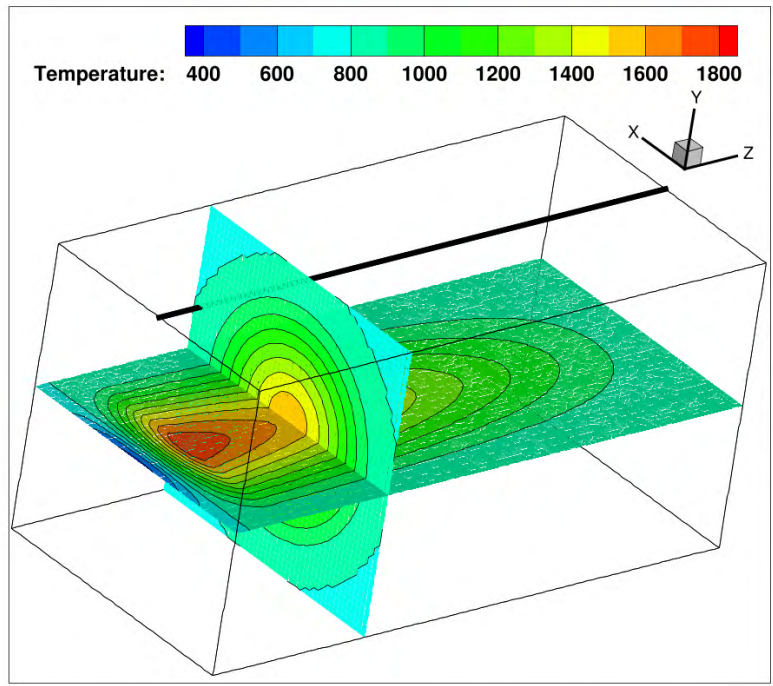

Figure 1: Contours of the temperature (in K) inside the box problem.

Table 1: Mixture Composition for Test Cases

| Case | Related Combustion Problem | vol%-$N_2$ | vol%-$H_2O$ | vol%-$CO_2$ |
|------|---------------------------|------------|-------------|-------------|
| 1 | air-methane | 70 | 20 | 10 |
| 2 | air-oil (possibly air-coal) | 80 | 10 | 10 |
| 3 | oxy-coal, wet recycle | 0 | 35 | 65 |
| 4 | oxy-coal, dry recycle | 0 | 10 | 90 |

## 2 Formulation

### 2.1 RTE and WSGGM

The gray-gas form of the RTE along a line of sight in an emitting/absorbing medium is [15, 16]

$$\frac{d\,I_{tot}}{d\,s} = k\left(\frac{\sigma T^4}{\pi} - I_{tot}\right) \tag{4}$$

where $I_{tot}$ is the total radiative intensity, $s$ is a spatial coordinate, $\sigma$ is the Stefan–Boltzmann constant, $T$ is the temperature, and $k$ is the radiative absorption coefficient. The latter is computed here indirectly using the WSGGM in two steps. First, the total emissivity is computed as a weighted sum of 'hypothetical' gray gases [17–20]

$$\epsilon = \sum_{i=1}^{I} a_i(T)\left(1 - exp\left[-k_i\,p\,L\right]\right) \tag{5a}$$

$$\text{where} \quad a_i(T) = \sum_{j=1}^{J} b_{ij}\,T^{j-1} \tag{5b}$$

where $p$ is the sum of partial pressures of water vapor ($P_w$) and carbon dioxide ($P_c$), $L$ is the mean pathlength, $k_i$ are the pressure absorption coefficients for the $I$ gray gases, $a_i$ are weight factors, and $b_{ij}$ are the coefficients for a polynomial of degree $J - 1$ in T. Second, the absorption coefficient is obtained using the Beer–Lambert law as

$$k = -\frac{1}{L}\ln\left(1 - \epsilon\right) \tag{6}$$

Each WSGGM has a set of coefficients($k_i$ and $b_{ij}$) at a finite number of $H_2O$ and $CO_2$ partial pressures. The number of gray gases ($I$) and the order of the polynomial ($J - 1$) also vary from one model to another. Because a WSGGM has $I$ absorbing gray gases and one (dummy) clear gas, $k_0$=0, it will be denoted here as $I + 1$ WSGGM.

### 2.2 Comments on the Gray-Gas Assumption

It is useful to provide justification for our use of gray (versus non-gray) radiation modeling, Equation 4, before we proceed in this study. This formulation is based on the assumption that the absorption coefficient is independent of the wavelength. It allows solving a single RTE instead of a large number of RTEs. Although this assumption is unrealistic for gases, there have been many industrial problems where this approximation provided acceptable results for the wall radiative flux [21]. In addition, neglecting the influence of the spectral gaseous radiation in combustion simulations results in overprediction of the radiative heat flux; this in turn decreases the predicted temperatures. On the other hand, not accounting for the turbulence-radiation interaction (TRI) has a counter effect; which is a reason for why computational studies that did not account for both (computationally-expensive) aspects of radiation modeling (i.e., TRI and non-grayness) were able to reach good agreement with measurements for flames [22]. The present study shows that predictions from the gray WSGGM can agree within approximately 25% to those from the non-gray and more-sophisticated spectral line-based WSGGM (SLW). Given the fact that in combustion simulations, radiation is only one of multiple physical/chemical phenomena that are modeled (e.g., turbulence and reactions), using the gray gas approach seems like a reasonable load balancing strategy. This approach is often used in the CFD of combustion including radiation [13]. Finally, for coal combustion, radiative properties of the coal can dominate the radiative transport over much of the domain reducing the error penalty from using a gray-gas approximation [12, 15, 16].

## 2.3 WSGGM for Arbitrary $CO_2$ and $H_2O$ Partial Pressures

As described in the previous section, the WSGGM coefficients (the temperature-polynomial coefficients and gray-gas pressure absorption coefficients) for each version of the WSGG model are available at finite compositions. In practice, it is unlikely that the operating condition, particularly at any particular point in the domain, will be at any one the few tabulated compositions. Towards this end, we have investigated two methods to calculate the model coefficients at arbitrary $CO_2$ and $H_2O$ partial pressures. We will define and use the following mole fraction ratios, also used by Hottel et al. [23], (by mole or partial pressure) to concisely describe the relative $H_2O$-$CO_2$ content of a gas mixture:

$$R \equiv \frac{H_2O}{CO_2}, \quad RR \equiv \frac{H_2O}{H_2O + CO_2} \tag{7}$$

Whereas the range of the first ratio, $R$, is unbounded (from 0 to $\infty$), $RR$ is bounded between 0 and 1. We denote the particular values of $R$ or $RR$ at which model coefficients were developed for a particular WSGGM as the *tabulated* ratios, and the corresponding coefficients as the *tabulated* coefficients. In the present study, either piecewise constant or piecewise linear interpolation of these coefficients is used for each of the WSGGM. When using piecewise-constant interpolation, we used the arithmetic average of two adjacent tabulated $RR$ values to define the interval boundaries. The model of Yin et al. includes a list of interval boundaries where the $R$ range (under the condition that $p > 0.5$ atm) is divided into the following intervals: 0-0.2, 0.2-0.4, 0.4-0.6, 0.6-0.9, 0.9-1.1, 1.1-2.5, and 2.5-$\infty$. The respective sets of tabulated coefficients used within these intervals are those corresponding to $R$ = 0.125, 0.25, 0.5, 0.75, 1, 2, and 4. For the piecewise-linear implementation, the tabulated composition ratios are used as the "nodes" for the linear interpolation. In the interval containing $RR$=0 or $RR$=1 where no tabulated coefficients are available at either end, extrapolation is applied using the closest two tabulated sets of coefficients. To illustrate both implementation techniques, the details of each one are given in Appendices A and B for the WSGGM of Smith et al., as an example.

## 3 Analyzed WSGG Models

The five WSGG models which we have studied are described in the following subsections.

## 3.1 The Johansson et al. 3+1 and 4+1 WSGGMs

Both models [6] were developed at Chalmers University of Technology in Sweden. They are based on emissivities calculated with the EM2C [24] SNB (statistical narrowband) model, and each one has tabulated coefficients sets at only 2 compositions.

1. Mixture of $H_2O$/$CO_2$, with mole fractions 0.1111/0.8889

2. Mixture of $H_2O$/$CO_2$, with mole fractions 0.5/0.5

The models are calibrated for a mixture (total) pressure of 1 bar (nearly atmospheric), a pressure weighted pathlength, $pL$, from approximately 0.01 atm-m to 60 atm-m, and a temperature range from 500 K-2500 K. The model polynomials are quadratic in temperature.

## 3.2 The Krishnamoorthy et al. 3+1 WSGGM

The model [7], a result of collaboration between ANSYS-FLUENT and the National Energy Technology Laboratory of the U.S. Department of Energy, is based on emissivities from the empirical correlation [1] in the Perry's Chemical Engineers' Handbook [23]. The model has tabulated coefficients for the following three compositions:

1. Mixture of $N_2/H_2O/CO_2$, with mole fractions 0.8/0.1/0.1
2. Mixture of $H_2O/CO_2$, with mole fractions 0.35/0.65
3. Mixture of $H_2O/CO_2$, with mole fractions 0.1/0.9

The Perry's Handbook correlation is based on based on data in classical Hottel emissivity charts [17, 20]. These data have been adjusted with more recent measurements in RADCAL [25]. The correlation is applicable to atmospheric total pressure, and has a stated range of applicability for a pressure-pathlength between 0.005 atm-m to 10 atm-m, and temperatures between 1000 K and 2000 K. The temperature polynomials are linear for this WSGGM.

## 3.3 The Yin et al. 4+1 WSGGM

The model [8] was developed at Aalborg University in Denmark. It is based on emissivities calculated from an exponential wide band model (EWBM) [26, 27]. The model provides coefficients sets for 10 different compositions.

1. $H_2O$ and $CO_2$, with limiting mole fractions $\rightarrow 0$
2. $H_2O$ and $CO_2$, with mole fractions 0.1 and 0.1
3. $H_2O$ and $CO_2$, with mole fractions 0.3 and 0.1
4. Mixture of $H_2O/CO_2$, with mole fractions 0.1111/0.8889
5. Mixture of $H_2O/CO_2$, with mole fractions 0.2/0.8
6. Mixture of $H_2O/CO_2$, with mole fractions 0.3333/0.6667
7. Mixture of $H_2O/CO_2$, with mole fractions 0.4286/0.5714
8. Mixture of $H_2O/CO_2$, with mole fractions 0.5/0.5
9. Mixture of $H_2O/CO_2$, with mole fractions 0.6667/0.3333
10. Mixture of $H_2O/CO_2$, with mole fractions 0.8/0.2

This model is derived for mixture pressure of 1 atm. The pressure-pathlength range is from 0.001 atm-m to 60 atm-m and the temperature range is 500 K-3000 K. The model polynomials are cubic in temperature.

## 3.4 The Smith et al. 3+1 WSGGM

The model [9] was developed at the University of Iowa. It is based on emissivities generated using an EWBM [28, 29]. Sets of tabulated coefficients are available for the following gas mixtures:

1. $CO_2$ with limiting partial pressure $P_c \rightarrow 0$.
2. $H_2O$ with limiting partial pressure $P_w \rightarrow 0$.
3. Pure $H_2O$.

---

[1]There are two correlation formulas: a simple one but with restricted applicability and a detailed one but more general. This WSGGM utilized the more general formula.

4. Mixture of $N_2/H_2O/CO_2$, with mole fractions 0.8/0.1/0.1 (representative of air-combustion of oil).

5. Mixture of $N_2/H_2O/CO_2$, with mole fractions 0.7/0.2/0.1 (representative of air-combustion of $CH_4$).

The total pressure is 1 atm. The range of $pL$ is from 0.001 atm-m to 10 atm-m. The temperature range for the model is 600 K-2400 K. The model polynomials are cubic in temperatures.

Table 2 summarizes and compares main characteristics of the WSGGMs we applied here.

Table 2: Key Features of Different WSGGMS

| Name and Year | Abb. | Gases | Tabulated Sets | $pL$ (in atm-m) | T (in K) | Polynomial |
|---|---|---|---|---|---|---|
| Johansson et al. 2010 | J31 | 3+1 | 2 | 0.01-60 | 500-2500 | Quadratic |
| Johansson et al. 2010 | J41 | 4+1 | 2 | 0.01-60 | 500-2500 | Quadratic |
| Krishnamoorthy et al. 2010 | K31 | 3+1 | 3 | 0.005-10 | 1000-2000 | Linear |
| Yin et al. 2010 | Y41 | 4+1 | 10 | 0.001-60 | 500-3000 | Cubic |
| Smith et al. 1982 | S31 | 3+1 | 5 | 0.001-10 | 600-2400 | Cubic |

# 4 Results

We will use abbreviations (see Table 2) in the plots to designate each WSGGM and the interpolation technique, such as (J31s) for Johansson et al. 3+1 WSGGM with piecewise-constant interpolation (the suffix 's' corresponds to 'stepwise'), and (Y41i) for Yin et al. 4+1 WSGGM with piecewise-linear interpolation (the suffix 'i' stands for 'interpolation'). An abbreviation without the suffix (s) or (i) designates a WSGGM applied in a test case that matches a gas composition for which the model provides a tabulated set of coefficients.

## 4.1 Effect of Gas Composition and Interpolation

We calculated the emissivity and absorption coefficient over $H_2O/CO_2$ ratios between 0 and 2, with the five WSGGMs using both interpolation methods. The variation with respect to the $H_2O/CO_2$ ratio was calculated for pathlengths of 1 m (Figure 2) and 10 m (Figure 3). The mixture in these profiles corresponds to $H_2O$-$CO_2$ with atmospheric sum of their partial pressures (no third species), which is a representation of oxy-fuel carbon capture combustion. The largest deviation between the classical and new models occurs at the lower $H_2O/CO_2$ ratios (below 0.35). This is the operating regime of oxy-fuel combustion with dry-recycle. Piecewise-linear interpolation reduces this difference in this range. As we will see in subsection 4.2, the increase in the absorption coefficient calculated from the classical model when linear interpolation is used significantly improves the predictions for low but finite $H_2O$ partial pressure. Also comparing Figure 2 and Figure 3, the difference is larger for the larger pathlength calculations. Therefore, one expects that these trends would also hold for boiler simulations, with a larger deviation for production systems than for lab or pilot scale systems. The figures also highlight the discontinuity in the properties for different intervals.

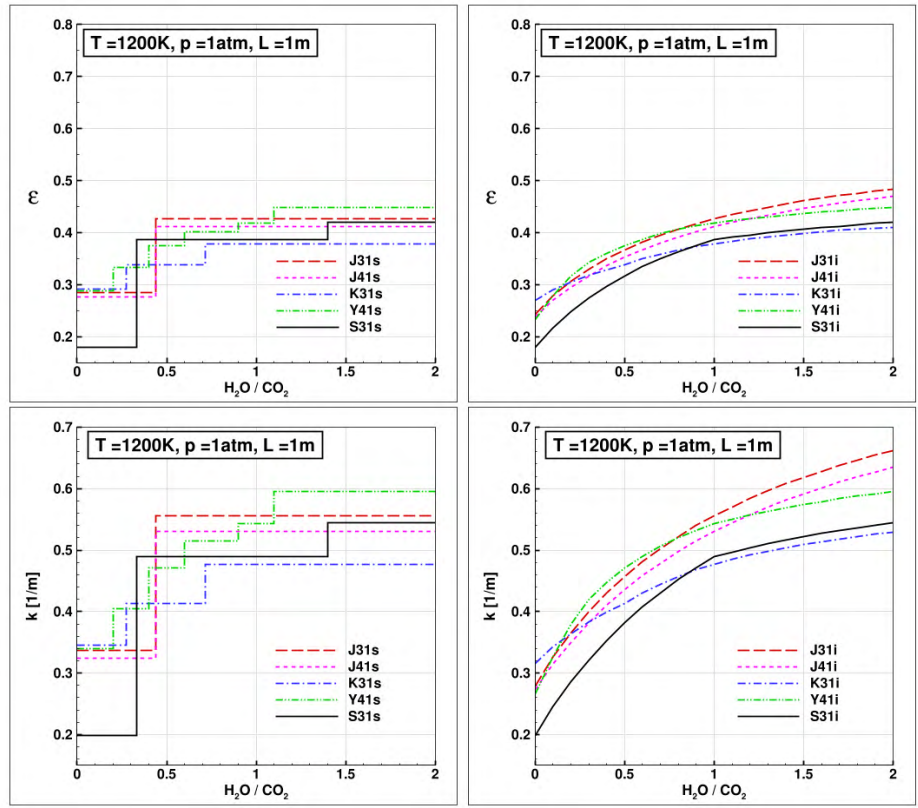

Figure 2: Profiles of emissivity and absorption coefficient for atmospheric $H_2O$-$CO_2$ mixtures with $L$=1 m.

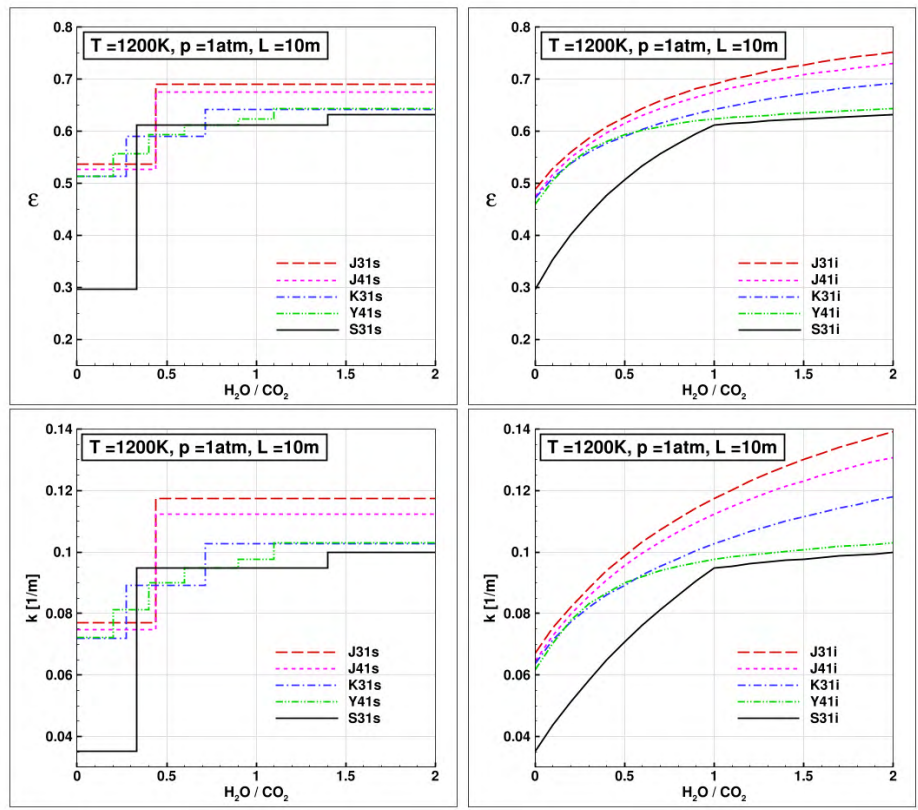

Figure 3: Profiles of emissivity and absorption coefficient for atmospheric $H_2O$-$CO_2$ mixtures with $L$=10 m.

## 4.2 Solution of the Radiative Equation of Transfer

The RTE is solved for the box furnace described in the introduction. Radiative properties are provided by the five WSGGM with the two interpolation methods. We implemented the models as user-defined functions (UDFs) for ANSYS FLUENT 12.1 [11] which was used to solve the RTE using the finite volume method. The mean path length, $L$, was approximated by 3.6 volume/area [15], which gives $L$=1.44 m. The equations are solved with a spatial resolution of 41×41×80 and an angular resolution of 7×7 (zenith × azimuthal angles) in each octant (392 total divisions of the entire $4\pi$ solid-angle space). Figure 4 shows the solution to the first test case with J31s WSGGM at finer (8×8) and coarser (6×6) angular resolutions (left) as well as a finer (50×50×100) spatial resolution (right). We assume that the resolution is adequate since the differences between the solutions are visually unnoticeable.

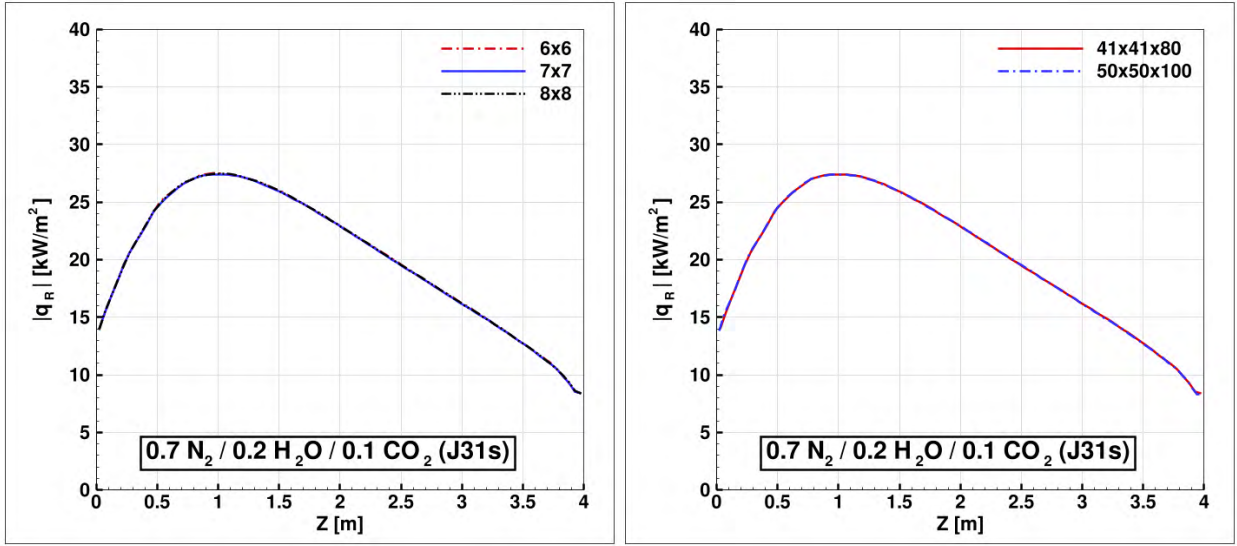

Figure 4: Illustration of the insensitivity of heat-flux prediction to angular and spatial resolution.

Figure 5 shows plots of the calculated magnitude of the radiative heat ?ux along the center line of upper side (Figure 1 - heavy solid line) using the different WSGGMs. Each row of plots in the figure corresponds to a different gas mixture, with a separate plot for each of the two interpolation methods. For the second test, all the models provide a tabulated set of coefficients at this molar ratio ($H_2O/CO_2 = 1$). Therefore, both interpolation methods calculate the same values of the model coefficients and thus only one set of curves is plotted. All plots also contain results using the SLW model. These are unpublished results generated by Dr. Gautham Krishnamoorthy. For the SLW solution, the absorption cross-section domains of $H_2O$ and $CO_2$ were individually discretized into 20 logarithmically spaced intervals between $3\times10^{-5}$ m$^2$/mol and 120 m$^2$/mol for $H_2O$, and between $3\times10^{-5}$ m$^2$/mol and 600 m$^2$/mol for $CO_2$. The analytical expressions for the absorption-line blackbody distribution functions (ALBDFs) of $H_2O$ [30] and $CO_2$ [31] were used to compute the blackbody weights of each gray gas. A reference temperature characterized by a representative temperature over the domain was selected, then the scaling approximation was used to relate the ALBDF at the reference state with that at the local temperature. The double integration method [32] was used to compute the blackbody weights and absorption coefficients for $H_2O$-$CO_2$ mixtures. The number of RTEs was 441 (one RTE per each of the 21×21 gray gases). The SLW calculations were performed using the T4 angular quadrature [33], and with a spatial resolution of 26×19×19 using a proprietary discrete ordinates radiation code.

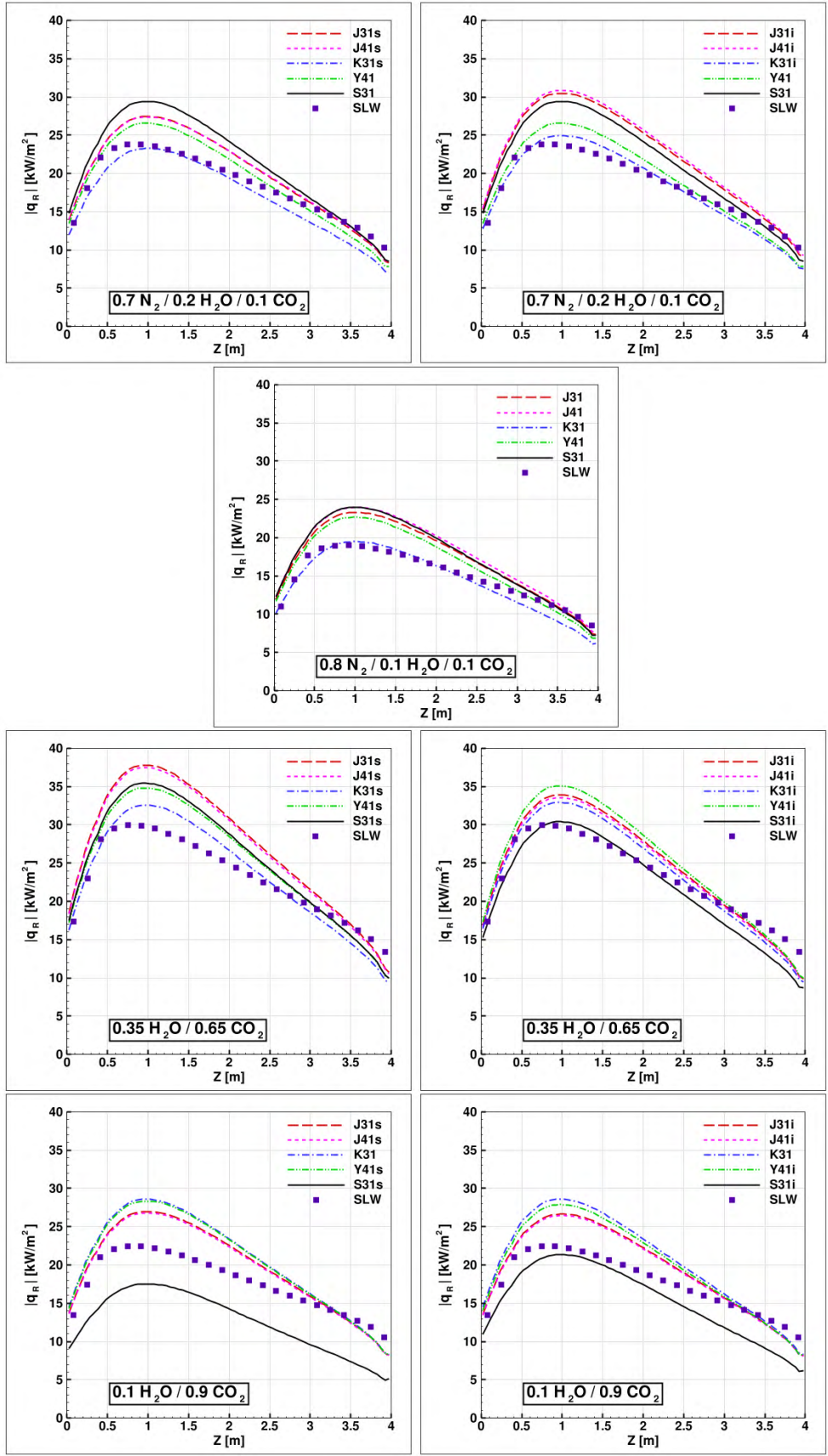

Figure 5: Comparisons of predicted radiative heat fluxes at different gas-mixture compositions.

Table 3: Estimates of the Difference between the WSGGMs and SLW Results for Test Case # 1

| N$_2$/H$_2$O/CO$_2$ = 0.7/0.2/0.1 | | | | | | | | |
|---|---|---|---|---|---|---|---|---|
| **WSGG−SLW** | J31s | J31i | J41s | J41i | K31s | K31i | Y41 | S31 |
| **Mean (kW/m$^2$)** | 1.73 | 3.92 | 1.77 | 4.24 | -1.38 | -0.20 | 0.80 | 2.87 |
| **RMS (kW/m$^2$)** | 2.38 | 4.51 | 2.37 | 4.81 | 1.66 | 1.22 | 1.81 | 3.61 |

Table 4: Estimates of the Difference between the WSGGMs and SLW Results for Test Case # 2

| N$_2$/H$_2$O/CO$_2$ = 0.8/0.1/0.1 | | | | | |
|---|---|---|---|---|---|
| **WSGG−SLW** | J31 | J41 | K31 | Y41 | S31 |
| **Mean (kW/m$^2$)** | 2.31 | 2.87 | -0.56 | 1.60 | 2.58 |
| **RMS (kW/m$^2$)** | 2.81 | 3.33 | 1.07 | 2.28 | 3.19 |

Table 5: Estimates of the Difference between the WSGGMs and SLW Results for Test Case # 3

| H$_2$O/CO$_2$ = 0.35/0.65 | | | | | | | | | | |
|---|---|---|---|---|---|---|---|---|---|---|
| **WSGG−SLW** | J31s | J31i | J41s | J41i | K31s | K31i | Y41s | Y41i | S31s | S31i |
| **Mean (kW/m$^2$)** | 4.27 | 1.57 | 3.98 | 1.32 | 0.48 | 0.73 | 2.08 | 2.26 | 2.39 | -1.22 |
| **RMS (kW/m$^2$)** | 5.21 | 2.67 | 4.95 | 2.48 | 2.02 | 2.16 | 3.19 | 3.35 | 3.54 | 2.05 |

Table 6: Estimates of the Difference between the WSGGMs and SLW Results for Test Case # 4

| H$_2$O/CO$_2$ = 0.1/0.9 | | | | | | | | | |
|---|---|---|---|---|---|---|---|---|---|
| **WSGG−SLW** | J31s | J31i | J41s | J41i | K31 | Y41s | Y41i | S31s | S31i |
| **Mean (kW/m$^2$)** | 2.23 | 1.99 | 2.07 | 1.83 | 3.07 | 2.92 | 2.59 | -5.10 | -2.31 |
| **RMS (kW/m$^2$)** | 2.99 | 2.76 | 2.84 | 2.63 | 3.94 | 3.79 | 3.48 | 5.13 | 2.58 |

The RMS and mean deviations between the WSGGMs solution and the benchmark solution from the SLW model are summarized in Tables 3-6 for the four test cases.

Based on these results we make the following points:

- The two WSGGMs of Johansson et al. give almost identical results. Therefore, reducing the number of gray gases from 4+1 to 3+1 did not affect the capability of the WSGGMs (but had a favorable effect of reducing the amount of computations).

- The Krishnamoorthy et al. WSGGM had the lowest overall deviation from the SLW model for the two air-fuel test cases. This is an interesting finding knowing that the model has the simplest (linear) temperature polynomial, and 3+1 (not 4+1) gray gases.

- None of the new models shows clear superiority when considering the two oxy-fuel test cases. In the fourth test case, with the highest concentration of CO$_2$ (moving towards pure CO$_2$), the disparity between the predictions of the new models is substantially reduced.

- The fourth test case (highest concentration of CO$_2$) shows considerable deviation of the piecewise-constant air-fuel WSSGM (Smith et al.) from the results of the SLW and new WSGG models, with underprediction of the radiative heat ?ux (we found a similar behavior for the radiative source term, calculated along the box centerline). However, the piecewise-linear interpolation has appreciably improved the prediction of this model.

- The piecewise-linear interpolation, in general, reduces the spread in the predictions of the various models as well as reducing the model deviation from the SLW predictions. The Smith et al. model for the oxy-fuel cases is the most sensitive to the coefficients interpolation approach, reducing differences between the model predictions and those from the SLW by about 50% in the two oxy-fuel test cases. The exceptions to this trend are the two Johansson et al. models for air-firing, where the results get worse not better. Which is not surprising since the models were developed for oxy-fuel combustion and do not provide sufficient tabulated coefficients to cover this particular $CO_2$ and $H_2O$ partial pressures.

# 5 Concluding Remarks

We have compared finite-volume solutions to the radiative equation of transfer (RTE) using the classical air-fuel WSGGM of Smith et al. (1982) and four new oxy-fuel WSGGMs (2010). In general the models overpredict the radiative heat flux relative to the SLW model for both oxy and air conditions. In the dry-recycle case, however, the classical model underpredicts the radiative heat flux. This is due to a calculated absorption coefficient which is significantly lower than that calculated by the other models in this operating regime. Linear interpolation of the model coefficients increases the absorption coefficient for low but finite $H_2O$ partial pressure and consequently reduces the deviation between the heat flux calculated from the classical WSGGM and the SLW model. For wet-recycle the heat flux using the classical model is within the range of the other models.

The limited test performed herein did not identify a specific model which was better than the others. Linear interpolation of the model coefficients generally improved the results. Overall, the WSGGM developed by Krishnamoorthy et al. had the smallest deviation from the SLW results of all the models considering all 4 test cases. This combined with its lower computational cost, due to fewer coefficients, make this an attractive model. However, The limited temperature range (1000 K-2000 K) and pressure-pathlength (10 atm-m) are disadvantages. The model of Yin et al. performed nearly as well as the model of Krishnamoorthy et al., however the former has a much wider operating range, 500 K-3000 K for temperature with pressure-pathlengths up to 60 atm-m. The two models of Johansson et al. do not provide sufficient coefficients for air fired conditions, so the higher error for these models in the air-fired cases is understandable. However, for the dry-recycle case, the 4 gray + 1 clear gas version of the model had the lowest deviation from the SLW results of all the models.

# 6 Acknowledgments

This technical effort was performed in support of the National Energy Technology Laboratory's ongoing research in $CO_2$ Capture under the RES contract DE-FE0004000. The thank Dr. Krishnamoorthy for the use of his SLW results.

# Appendix A: Piecewise-Constant Implementation of Smith et al. WSGGM

Table 7: Coefficients and Piecewise-Constant Implementation for Smith et al. Model

| gray gas index: $i$ | $k_i$ | $b_{i1}$ | $b_{i2}$ | $b_{i3}$ | $b_{i4}$ |
|---|---|---|---|---|---|
| For $RR \leq 1/4$[a] ($R \leq 1/3$) Use coefficients for $RR \to 0$ ($R \to 0$): $P_c \to 0$, no $H_2O$ | | | | | |
| 1 | 0.3966 | 0.4334e-1 | 2.620e-4 | -1.560e-7 | 2.565e-11 |
| 2 | 15.64 | -0.4814e-1 | 2.822e-4 | -1.794e-7 | 3.274e-11 |
| 3 | 394.3 | 0.5492e-1 | 0.1087e-4 | -0.3500e-7 | 0.9123e-11 |
| For $RR \leq 7/12 = 0.58333$[b] ($R \leq 1.4$) Use coefficients for $RR = 1/2$ ($R = 1$) | | | | | |
| 1 | 0.4303 | 5.150e-1 | -2.303e-4 | 0.9779e-7 | -1.494e-11 |
| 2 | 7.055 | 0.7749e-1 | 3.399e-4 | -2.297e-7 | 3.770e-11 |
| 3 | 178.1 | 1.907e-1 | -1.824e-4 | 0.5608e-7 | -0.5122e-11 |
| For $RR \leq 5/6 = 0.83333$[c] ($R \leq 5$) Use coefficients for $RR = 2/3$ ($R = 2$) | | | | | |
| 1 | 0.4201 | 6.508e-1 | -5.551e-4 | 3.029e-7 | -5.353e-11 |
| 2 | 6.516 | -0.2504e-1 | 6.112e-4 | -3.882e-7 | 6.528e-11 |
| 3 | 131.9 | 2.718e-1 | -3.118e-4 | 1.221e-7 | -1.612e-11 |
| For $RR > 5/6$ ($R > 5$) AND $P_w \leq 0.5$ atm Use coefficients for $RR \to 1$ ($R \to \infty$): $P_w \to 0$, no $CO_2$ | | | | | |
| 1 | 0.4098 | 5.977e-1 | -5.119e-4 | 3.042e-7 | -5.564e-11 |
| 2 | 6.325 | 0.5677e-1 | 3.333e-4 | -1.967e-7 | 2.718e-11 |
| 3 | 120.5 | 1.800e-1 | -2.334e-4 | 1.008e-7 | -1.454e-11 |
| For $RR > 5/6$ ($R > 5$) AND $P_w > 0.5$ atm Use coefficients for $RR = 1$ ($R = \infty$): pure $H_2O$ at 1 atm | | | | | |
| 1 | 0.4496 | 6.324e-1 | -8.358e-4 | 6.135e-7 | -13.03e-11 |
| 2 | 7.113 | -0.2016e-1 | 7.145e-4 | -5.212e-7 | 9.868e-11 |
| 3 | 119.7 | 3.500e-1 | -5.040e-4 | 2.425e-7 | -3.888e-11 |

[a] We used the arithmetic mean of 0 and 1/2,    [b] the arithmetic mean of 1/2 and 2/3,    [c] the arithmetic mean of 2/3 and 1.

# Appendix B: Piecewise-Linear Implementation of Smith et al. WSGGM

IF $RR \leq 1/2$ :
    Linearly interpolate over $RR$ using the tabulated coefficients at $RR = 0$ & $1/2$ ($R = 0$ & $1$). For example,

$$\text{let } \delta = \frac{RR - 0}{1/2 - 0}$$

$$\text{then } k_i = k_i(RR = 0) + \delta \ [k_i(RR = 1/2) - k_i(RR = 0)]$$

$$\text{and } b_{ij} = b_{ij}(RR = 0) + \delta \ [b_{ij}(RR = 1/2) - b_{ij}(RR = 0)]$$

ELSE IF $RR \leq 2/3$:
    Linearly interpolate over $RR$ using the tabulated coefficients at $RR = 1/2$ & $2/3$ ($R = 1$ & 2)

ELSE:
    Linearly interpolate over $RR$ using the tabulated coefficients at $RR = 2/3$ & 1 ($R = 2$ & $\infty$)

    if $P_w \leq 0.5$ atm:
        Use the tabulated coefficients for ($P_w \to 0$, no $CO_2$) to represent the $RR = 1$ condition

    else:
        Use the tabulated coefficients for (pure $H_2O$ at 1 atm) to represent the $RR = 1$ condition